\documentclass[aps,twocolumn,showpacs,amsmath,amssymb]{revtex4}

\usepackage{graphicx}

\newcommand{\eone}{\hat{E}_1}
\newcommand{\eoned}{\hat{E}_1^{\dagger}}

\newcommand{\wone}{\hat{\Omega}_1}

\newcommand{\woned}{\hat{\Omega}_1^{\dagger}}

\newcommand{\etwo}{\hat{E}_2}
\newcommand{\etwod}{\hat{E}_2^{\dagger}}

\newcommand{\be}{\begin{eqnarray}}
\newcommand{\ee}{\end{eqnarray}}

\begin{document} 
\title{Efficient photon counting and single-photon generation using resonant
nonlinear optics}
\author{Mattias Johnsson and Michael Fleischhauer} 
\affiliation{Fachbereich Physik, Universit\"{a}t Kaiserslautern, D-67663 Kaiserslautern, 
Germany} 
\date{\today} 
 
\begin{abstract} 
The behavior of an atomic double lambda system in the presence of a
strong off-resonant classical field and a few-photon resonant quantum field is
examined. It is shown that the system possesses properties that allow
a single-photon state to be distilled from a multi-photon input
wave packet. In addition, the system is also capable of functioning as
an efficient photodetector discriminating between one- and two-photon
wave packets with arbitrarily high efficiency.
\end{abstract} 
 
\pacs{42.50.Gy, 42.50.Md, 42.50.Ar} 
 
\maketitle

Recently Knill, Laflamme, and Milburn (KLM) \cite{KnillNature2001} have 
proposed a probabilistic scheme for efficient quantum computation 
using only linear optical elements, sources of entangled photons and 
efficient photodetectors. This scheme, based on Gottesman
and Chuang's discovery that universal quantum computation requires
only teleportation and single qubit operations \cite{GottesmanNature1999},
has attracted much attention since it provides a very interesting alternative
to schemes based on controlled nonlinear qubit-qubit interactions, which are
extremely difficult to implement on a larger scale. 
One of the practical challenges of the KLM scheme, however, is 
the requirement of photodetectors that can distinguish between zero, 
one and two photons. State of the art photon counters have 
rather small quantum efficiencies, and high efficiency avalanche 
detectors cannot discriminate between one
and two photons. Another basic requirement of the KLM proposal, as 
in any photon-based quantum computation scheme, is the ability to 
generate single-photon wave packets in a controlled manner. In this paper
we show that both tasks may be solved by making use of resonant nonlinear
optical processes based on electromagnetically induced transparency (EIT)
\cite{harris1997,Marangos}. 

EIT and similar interference effects have led to a new regime 
of nonlinear optics on the level of few light quanta 
\cite{Harris1998,HauHarris99,Imamoglu96,Lukin-AdvAMO-2000,johnsson2002a}
as they allow making use of the resonantly enhanced nonlinear 
interaction without suffering from linear absorption and refraction. 
The potential for strong interactions between individual photons makes
EIT-based nonlinear optics a promising candidate for the
implementation of quantum gate operations. However, although it has
been shown that resonant nonlinear 
interactions are strong enough to obtain, for example, a 
conditional phase shift
of one photon by another \cite{Imamoglu96}, so far no scheme exists that
allows a {\it controlled} qubit-qubit interaction for photon wave packets.
The situation is different, however, if only one of the
interacting fields is a pulse, as is the case in our scheme.

In this paper we show that resonant four-wave mixing in an atomic double
lambda configuration \cite{4WM} with one strong coherent input can be used
to filter out a single-photon wave packet from a given input. The same filtering
technique can be used to select components of an incoming wave packet according
to their photon number and direct them to different high efficiency 
avalanche photodetectors. In this way efficient photodetectors can be built 
that can distinguish between zero, one and two photons.

\begin{figure}[ht] 
  \begin{center} 
    \includegraphics[width=8cm]{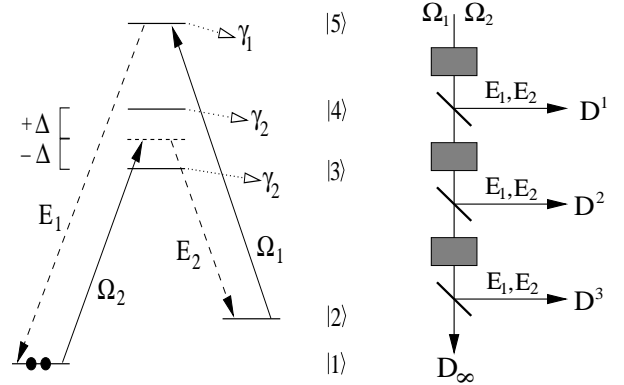} 
    \caption{{\it left:} Four-wave mixing in a 
modified double-$\Lambda$ system
with sgn[$d_{42}/d_{41}]=-$sgn$[d_{32}/d_{31}]$, with $d_{ij}$ being the
dipole moment of the $|i\rangle - |j\rangle$ transition.
{\it right:} Cascade detection system consisting of repeated sections of 
nonlinear medium and beamsplitter. In each section the $\Omega_1$
field is partially converted to the $E_1$ and $E_2$ fields, which are
then diverted by the beamsplitter for filtering or detection. 
}
    \label{fig5level} 
  \end{center} 
\end{figure}

We consider resonantly enhanced forward four-wave mixing in the
modified double-lambda system shown in Fig.~\ref{fig5level}
where nonlinear phase shifts are
eliminated by AC-Stark compensation \cite{johnsson2002b}. 


\noindent Here
two fields with slowly varying amplitudes 
$\Omega_1$ and $\Omega_2$ are initially excited and form
the pump fields. The other fields with slowly varying amplitudes
$E_1$ and $E_2$ are assumed to be initially zero. 
$\Omega_1$ and $E_1$ are taken to be exactly on resonance,
while the other two fields are detuned by an amount
$\Delta$. Decay from the two lower levels is considered to be
negligible and all fields have the same propagation direction. 
Because of energy conservation there is overall
four-photon resonance. It can also be shown  
that phase matching will favor pairwise  
two-photon resonance with the $|1\rangle-|2\rangle$ transition. 

Semiclassical solutions to this system
show that there is a cycling of energy between the
pump and generated fields with unit efficiency
\cite{4WM}. In general quantum 
effects prevent perfect conversion, but conversion efficiency of unity or
close to unity is possible in certain cases.
Due to the resonant nature of the system, the distance required
for a complete cycle becomes shorter as
the input power is reduced, which is in sharp contrast to 
ordinary off-resonant four-wave mixing \cite{johnsson2002a}.

This dependence of conversion length on input power raises the
possibility of using the double-lambda system for the creation of
single-photon states: One chooses the length of the nonlinear medium
such that a single-photon input state performs exactly one cycle to
the generated fields and back again, i.e. the output state at the exit
of the medium is the same as at the input, at least up to a phase.
Inputs consisting of higher photon numbers, however, will undergo
only a partial cycle, resulting in a lower chance of finding photons
in the original high photon number state at the exit of the medium. If 
correctly filtered after each passage, multiple stages successively
reduce higher photon number states, until eventually all that is left
is a single photon state and vacuum. As we will show, the process of filtering
also allows for the use of such a cascade system as an effective
detector discriminating between single and double photons. 

We assume that the drive field $\Omega_2$ is strong, and treat it 
classically, while the other three fields are treated fully 
quantum mechanically.
We use the setup shown in Fig.~\ref{fig5level}, which
consists of repeated stages of a length of nonlinear medium followed
by a beamsplitter. 
Three stages 
are shown, but in principle any number
can be used. At the input we assume that the strong classical field,
$\Omega_2$, is present, as is the few-photon quantum wave packet
described by the operator $\hat\Omega_1$. The generated fields $\hat E_1$ 
and $\hat E_2$ are assumed 
to be in the vacuum state at the input. The detectors $D^i$ are
standard, high-efficiency avalanche
photodetectors that will register the presence of a photon with
near certainty, but are unable to discriminate between single and
multiple photons. The beamsplitters are considered transmitting for
the two fields $\hat\Omega_1$ and $\Omega_2$ and reflective for the fields
$\hat E_1$ and $\hat E_2$. This could be accomplished for example by choosing 
orthogonal linear polarization for the $\Omega_j$ and $E_j$ fields and using
polarizing beam splitters.

The effective multi-mode Hamiltonian describing the interaction between
the fields in the presence of the nonlinear medium is given by
\cite{johnsson2002a}
\begin{equation}
\hat H_{\rm int} = \frac{\hbar g c}{\Delta} \int {\rm d}z 
\frac{\woned\Omega_2^* \eone \etwo + \eoned\etwod \wone\Omega_2}{\woned\wone
+\eoned\eone}
\label{eqHint}
\end{equation}
Here $\hat\Omega_1(z), \hat{\Omega}_1^{\dagger}(z)$ 
etc. denote dimensionless, slowly-varying (both in time and space) 
positive and negative-frequency components of the corresponding
electric field operators.
The three quantum fields $\hat\Omega_1$, $\hat E_1$, and $\hat E_2$ 
are taken to be independent and thus commute. The commutator 
between operators $\hat O$ and $\hat O^\dagger$ corresponding to
the same field is approximately a spatial delta function.
The coupling constant $g$ in (\ref{eqHint}) 
is given by $g= 3 N \lambda^2 \gamma/8 \pi$, where $N$ is 
the atomic number density,
$\lambda$ some typical wavelength of the fields and $\gamma$ the 
typical radiative decay rate. Since the numerator and denominator in 
(\ref{eqHint}) commute there is no ambiguity with respect to ordering.
The difference between the resonant four-wave mixing process and ordinary
off-resonant systems is in the unusual denominator of (\ref{eqHint}).
It results from the saturation of the
two-photon transition $|1\rangle - |2\rangle$ by the resonant fields
$\hat \Omega_1$ and $\hat E_1$.
Due to the saturation denominator the atomic system
has the largest nonlinear response when $\hat \Omega_1$ and $\hat E_1$
are small. 

It can be shown that the operator equations of motion are given by
 \cite{johnsson2003a}
\begin{equation}
\bigl(\partial_t + c\partial_z\bigr)\hat{\Omega}_1(z,t) = \frac{i}{\hbar} 
    \bigl[\hat H_{\rm int},\hat{\Omega}_1(z,t)\bigr]
\label{eqGeneralEOM}
\end{equation}
and similarly for $\hat{E}_1$ and $\hat{E}_2$. In the following
it is convenient to
introduce co-moving coordinates $z \rightarrow  \zeta = z - ct$ and
$t \rightarrow  \tau = t$, so that $\partial_t +c\partial_z 
\rightarrow \partial_{\tau}$. 
The effective Hamiltonian has the following three 
independent constants of motion:
\begin{eqnarray}
\partial_{\tau}(\hat \Omega_1^\dagger \hat \Omega_1 + \hat E_1^\dagger\hat E_1) &=&
0 \nonumber \label{eqCOM1} \\
\partial_{\tau}(\hat{E}_1^{\dagger} \hat{E}_1 - \hat E_2^\dagger\hat E_2) &=&
 0 \nonumber \label{eqCOM2} \\
\partial_{\tau} (\hat \Omega_1^\dagger
\Omega_2^{*} \hat E_1 \hat E_2  + \hat \Omega_1 \Omega_2
\hat E_1^{\dagger} \hat E_2^{\dagger}) &=& 0. \label{eqCOM4}
\end{eqnarray}
These equations indicate that whenever the $\hat \Omega_1$ field loses a
photon, both $\hat E_1$ and $\hat E_2$ must gain one. In addition, the photon
number of fields $\hat E_1$ and $\hat E_2$ are perfectly correlated --- they
will always lose or gain a photon together. 

The interaction of few-photon pulses with the double-lambda medium
can most easily be described in terms of state vectors in
the co-moving frame $(\zeta,\tau)$. 
\be
i\hbar\,\partial_\tau |\psi(\tau)\rangle = \hat H_{\rm int} \, |\psi(\tau)\rangle.\label{SE}
\ee
The local character of the interaction (\ref{eqHint}) and the constants
of motion (\ref{eqCOM4}) allow a reduction of the multi-mode problem to
a small number of states depending on the total number of photons. 
For example, if we consider an initial
single-photon wave packet in $\hat\Omega_1$ and vacuum in $\hat E_1$
and $\hat E_2$, then the initial state is given by
\be
|\psi(\tau=0)\rangle=|\psi_{0}^{(1)}
\rangle \sim  \int \!\!{\rm d}\zeta\, f(\zeta)\, \hat\Omega_1^\dagger(\zeta)\, 
|0\rangle \equiv |1,0,0\rangle,
\ee
where $f(\zeta)$ characterizes the shape of the wave packet and is called
the single-photon wave-function. The interaction Hamiltonian
couples $|\psi_{0}^{(1)}\rangle$ to only one other state, namely
\be
|\psi_1^{(1)}\rangle \sim \int \!\!{\rm d}\zeta\, f(\zeta)
\hat E_1^\dagger(\zeta)\hat E_2^\dagger(\zeta)\, |0\rangle
\equiv |0,1,1\rangle
\ee
which represents a two-photon wave packet in the two generated fields
and in turn is coupled only back to  $|\psi_{0}^{(1)}\rangle$.
Thus for the given input the multi-mode problem can be mapped onto
a two-state one. We have labeled the two states 
as $|1,0,0\rangle$ and $|0,1,1\rangle$, which denote
the total number of photons in $\hat\Omega_1$, $\hat E_1$, and $\hat
E_2$. 

Similarly, if the input is a two-photon wave packet in $\hat\Omega_1$
\be 
|\psi(\tau=0)\rangle=|\psi_{0}^{(2)}\rangle &\sim & \iint\!\! {\rm d}\zeta{\rm d}\zeta^\prime\, 
f_2(\zeta,\zeta^\prime)\, \hat\Omega_1^\dagger(\zeta)\
\hat\Omega_1^\dagger(\zeta^\prime)
\, |0\rangle \nonumber\\
&\equiv& |2,0,0\rangle
\ee
the interaction Hamiltonian couples it only to the states
\be
|\psi_{1}^{(2)}\rangle &\sim & \iint\!\! {\rm d}\zeta{\rm d}\zeta^\prime\, 
f_2(\zeta,\zeta^\prime)\, \hat E_1^\dagger(\zeta)\hat E_2^\dagger(\zeta)
\hat\Omega_1^\dagger(\zeta^\prime)\, |0\rangle\nonumber\\
&\equiv & |1,1,1\rangle,\qquad{\rm and}\\
|\psi_{2}^{(2)}\rangle &\sim & \iint\!\! {\rm d}\zeta{\rm d}\zeta^\prime\,  
f_2(\zeta,\zeta^\prime)\, \hat E_1^\dagger(\zeta)\hat E_2^\dagger(\zeta)
\hat E_1^\dagger(\zeta^\prime)\hat E_2^\dagger(\zeta^\prime)
\, |0\rangle\nonumber\\
&\equiv & |0,2,2\rangle.
\ee

We thus can now introduce a simple notation 
to label all relevant states of the
system. If we assume that there are
initially $n$ photons in field $\hat\Omega_1$ and vacuum in $\hat E_1$ and
$\hat E_2$ then, due to the constants of motion, we see that we can choose
a basis for the radiation field that
has the form $|n-j, \, j, \, 
j\rangle$, i.e. a basis
indicating how many photons are in each field at any one time.

Let us consider how the system can be used as a photon filter. To do this we
need to know what happens to single- and multi-photon input packets
as they pass through each stage of nonlinear medium and beamsplitter. 
We first consider the case where a single-photon wave packet is
injected into the system in the field $\hat \Omega_1$. 
To determine the evolution of the wave packet in the nonlinear medium
we solve the Schr\"{o}dinger equation (\ref{SE}) and apply the initial
condition that $|\psi(0)\rangle = |1,0,0\rangle$. 
We find the solution to be
\begin{equation}
|\psi(\tau) \rangle = \cos [\kappa |\Omega_2| c\tau] 
\,|1,0,0\rangle - i \sin
[\kappa |\Omega_2|c\tau] \,|0,1,1\rangle
\end{equation}
where $\kappa =g/\Delta$ showing that the pulse cycles smoothly between the pump and generated
fields. If we choose the length of the nonlinear medium to be
a multiple of $L_0\equiv \pi/(\kappa |\Omega_2|)$ we see that on exiting 
the medium the pulse is once
again entirely in the $\Omega_1$ field, with no component of the $E_1$
and $E_2$ fields excited. The shape of the pulse has an identical
shape to that of the input, but has undergone a phase shift of multiples of 
$\pi$.

\begin{figure}[ht] 
  \begin{center} 
    \includegraphics[width=8.5cm,height=4cm]{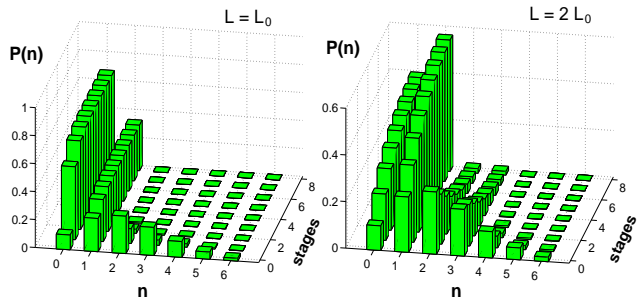} 
    \caption{Transfer of a coherent state input ($\langle n\rangle = 2.25$) 
into a single-photon wave packet over repeated stages
of nonlinear medium and filtering out of the $\hat E_1/\hat E_2$
components. For a minimum medium length of $L=L_0$ (left) 
the filtering process is faster but a larger vacuum component
survives compared to the case where $L=2L_0$ (right).
}
    \label{figFilterEfficiency} 
  \end{center} 
\end{figure}

Now consider the effect of the nonlinear medium on an input pulse
including higher photon numbers. Due to the constants of motion,
each distinct number state of the initial wave packet will evolve
separately in its own subspace. Thus, for example, the $n$-photon
component of the initial state can be considered to evolve distinctly
from the $n-1$ component. The action of the nonlinear medium on an
incoming $n$-photon state and the subsequent elimination of photons
in the $\hat E_1/\hat E_2$ modes by the beamsplitter is given by
\begin{equation}
|n , 0 , 0 \,\rangle \rightarrow \sum_{j=0}^n {\xi}^{(n)}_j |n-j, j , j \rangle
\rightarrow \sum_{j=0}^n {\bar \xi}^{(n)}_j |n-j, 0 , 0\rangle.
\end{equation}
The overbar denotes the fact that the coefficient may have
changed phase after the beamsplitter, but its absolute value is unchanged. 
As the $\xi_j$ all have magnitudes less than one, it is clear that each 
stage reduces the chance of 
finding the $\Omega_1$
field in photon occupation modes higher than one. 
Consequently the system converges on 
some combination of the $|1,0,0\rangle$ state and the vacuum state,
both of which are unaffected by the optical
elements. The system therefore serves as a method for generating
single-photon wave packets. As an example of the efficiency of the scheme,
in Figure~\ref{figFilterEfficiency} we present numerical results
for a coherent input and choose interaction lengths of $L_0$
and $2L_0$.

%
%

As a second possible application, we now demonstrate that 
the system also serves as a detector capable of discriminating between one and
two photons with high efficiency. To do this we need to consider the
dynamics of the system more carefully, taking the state of the
detectors placed at the exit ports $D^j$ into account and 
analytically finding values for the $\xi_j$ coefficients.

We indicate the state of the detector subsystem by
$|D^{i,j,k\ldots}\rangle$, signifying which detectors have fired. For
example $D^0$ indicates that no detectors have fired, $D^{1,3}$ would
indicate that detectors $D^1$ and $D^3$ have fired and so on.

First consider a single-photon pulse $|1 , 0 ,0\rangle$, which is sent
into the system with interaction segments of length $L_0$. 
After each traversal of a section of
nonlinear medium it will be in that same state, with a phase shift of
$\pi$. The
beamsplitter also leaves it in the same state, again albeit
with a possible change in phase. This is repeated for each stage in
the apparatus, until it reaches the $D^\infty$ detector. Thus,
if a single-photon wave packet is injected
into the system, none of the detectors put at the exits $D^1$ to $D^n$
will fire, while
the $D^\infty$ detector will fire with certainty. This, then, is the
signature of a single photon pulse.

Let us now consider the case where a two-photon pulse is injected into
the system. The action of the first nonlinear medium and the subsequent
beamsplitter on the input state is given by
\be
&&|2 \, 0 \, 0\rangle \rightarrow \xi^{(2)}_0 |2 \, 0 \, 0\rangle + \xi^{(2)}_1 |1
\, 1 \, 1\rangle + \xi^{(2)}_2 |0 \, 2 \, 2\rangle\label{eq2_photon_beamsplitter}\\
&&\quad\rightarrow \bar{\xi}^{(2)}_0 |2 \, 0 \, 0\rangle |D^0\rangle  + \bar{\xi}^{(2)}_1
|1 \, 0 \, 0\rangle |D^1\rangle + \bar{\xi}^{(2)}_2 |0 \, 0 \, 0\rangle |D^1\rangle
\nonumber
\ee
with the overbar indicating that some phase shift may have taken
place, i.e. $|\bar{\xi}^{(2)}_1| = |\xi^{(2)}_1|$, $|\bar{\xi}^{(2)}_2| = |\xi^{(2)}_2|$. If
we once again assume the length of the nonlinear medium is such that a
single photon state will undergo a simple sign change, after $n$
stages of nonlinear medium and beamsplitter we find that the state of
the input pulse reads 
\begin{eqnarray}
\xi^{(2)n}_0 |2 \, 0 \, 0 \rangle |D^0\rangle &+&
\sum_{k=0}^{n-1} (-1)^{k+n} \xi^{(2)k}_0 \bar{\xi}^{(2)}_1 |D^{k+1}\rangle
|1 \, 0 \, 0 \rangle  \nonumber \\
&+ & \sum_{k=0}^{n-1} \xi^{(2)k}_0 \bar{\xi}^{(2)}_2 |D^{k+1}\rangle
|0 \, 0 \, 0 \rangle \label{eq_psi_n_two_photon}.
\end{eqnarray}
This shows that three results are possible: $D^{\infty}$
fires and none of the $D^i$s does; $D^{\infty}$ fires and one of the
$D^i$ fires; one of the $D^i$ fires and $D^\infty$ does not.
The probability of the $i$th detector $D^i$ firing is 
$P(D^i) = |\xi^{(2)}_0|^{2i-2} -|\xi^{(2)}_0|^{2i}$. 
Crucially, the chance that {\em only}
$D^\infty$ will fire, thus giving a result indistinguishable from the
single-photon case, is given by $|\xi^{(2)}_0|^{2n}$. 

In order to obtain quantitative results we need to evaluate the
coefficients $\xi^{(2)}_0$, $\xi^{(2)}_1$ and $\xi^{(2)}_2$. The
$3\times 3$ equations of motion for the two-photon case 
can easily be integrated and yield
\begin{eqnarray}
\xi^{(2)}_0 &=& \frac{1}{3} \left( 2+\cos\sqrt{\frac{3}{2}}\kappa |\Omega_2|
c\tau \right) \nonumber \\
\xi^{(2)}_1 &=& -\frac{i}{\sqrt{3}}\sqrt{\frac{\Omega_2}{\Omega_2^*}}
\sin\sqrt{\frac{3}{2}}\kappa |\Omega_2| c\tau\\ 
\xi^{(2)}_2 &=& -2\frac{\sqrt{2}}{3}\frac{\Omega_2}{\Omega_2^*}
\sin^2\frac{1}{2}\sqrt{\frac{3}{2}}\kappa |\Omega_2| c\tau \nonumber
\end{eqnarray}
As we have assumed the length of the nonlinear medium to be
$c\tau=L_0 = \pi/\kappa |\Omega_2|$, we obtain the following numerical values
for the coefficients: $|\xi^{(2)}_0| = 0.4130$, $|\xi^{(2)}_1| = 0.3746$, and
$|\xi^{(2)}_2| = 0.8301$. Inserting these values 
into (\ref{eq_psi_n_two_photon}) 
we can calculate the state of the wave packet after each stage,
and the associated probability of any particular detector firing.
It is immediately clear that this scheme can
distinguish between one and two photons with great accuracy. For
example, if four stages are used, the chance of a two-photon
wave packet causing only the $D^{\infty}$ detector to fire is $|0.4130|^8=
8\times 10^{-3}$, i.e. one can distinguish between a single-photon wave packet
and a two-photon wave packet with greater than 99.9\% accuracy.
Using similar arguments one can show that the probability of only $D^\infty$
firing if the input pulse contains more than two photons is even smaller.
Thus the cascaded resonant nonlinear system 
when combined with avalanche detectors
can also be used as an efficient photodetector able to discriminate
between zero, one and many photons. 

It is worth mentioning that a set-up can be built along similar lines
in which only the two-photon or any other fixed 
photon-number component makes a full return after each interaction
zone. In this way a photon source or a detector can be built that
is tuned for example to a two- or three-photon wavepacket.

In summary we have shown that sucessive stages of 
resonant four wave mixing with one strong coherent cw input along with 
the filtering out of the generated fields can be used to transform
any low-photon input wavepacket into a single-photon wavepacket
with large probability. A similar set-up combined with
avalanche photodiodes can be used to build detectors that 
can discriminate between zero, one, and two or more photons.
Due to the resonant enhancement the required interaction lengths
are rather small and can be in the cm range \cite{johnsson2002a}.

The authors thank the Deutsche Forschungsgemeinschaft for financial support.

\def\etal{\textit{et al.}}

\vspace{.5cm}

\end{document}